\documentclass[preprint]{aastex}

\newcommand{\hMpc}{h^{-1}{\rm Mpc}}
\newcommand{\hGpc}{h^{-1}{\rm Gpc}}
\newcommand{\rmd}{{\rm d}}
\newcommand{\bx}{{\mathbf{x}}}
\newcommand{\by}{{\mathbf{y}}}
\newcommand{\bz}{{\mathbf{z}}}
\newcommand{\CB}{{\cal B}}
\newcommand{\CW}{{\cal W}}

\begin{document}

\title{A comparison of estimators for the two--point correlation function}
\author{Martin Kerscher\altaffilmark{1,2}, Istvan Szapudi\altaffilmark{3}, 
Alexander S. Szalay\altaffilmark{1}}
\altaffiltext{1}{Department of Physics and Astronomy, 
The Johns Hopkins University, Baltimore, MD 21218}
\altaffiltext{2}{Ludwig--Maximilians--Universit\"at, Theresienstra{\ss}e 37,
80333 M\"unchen, Germany, email: kerscher@theorie.physik.uni-muenchen.de}
\altaffiltext{3}{Canadian Institute for Theoretical Astrophysics,
University of Toronto, 60 St. George Street Toronto, Ontario, M5S 3H8, Canada}
\shorttitle{A comparison of estimators for $\xi(r)$}
\shortauthors{Kerscher, Szapudi \& Szalay}

\begin{abstract}
  Nine  of the  most  important estimators  known  for the  two--point
  correlation  function are compared  using a  predetermined, rigorous
  criterion.  The  indicators were extracted from  over 500 subsamples
  of the Virgo Hubble Volume simulation cluster catalog.  The ``real''
  correlation  function  was determined  from  the  full  survey in  a
  $3000\hMpc$  periodic  cube.   The  estimators were  ranked  by  the
  cumulative  probability  of  returning  a  value  within  a  certain
  tolerance of  the real  correlation function.  This  criterion takes
  into  account  bias and  variance,  and  it  is independent  of  the
  possibly non--Gaussian nature of  the error statistics.  As a result
  for astrophysical  applications a clear  recommendation has emerged:
  the {}\citet{landy:bias} estimator, in  its original or grid version
  {}\citep{szapudi:new},  are  preferred in  comparison  to the  other
  indicators  examined, with  a  performance almost  indistinguishable
  from the {}\citet{hamilton:towards} estimator.
\end{abstract}

\section{Introduction}

The two--point correlation function of galaxies became one of the most
popular statistical tools in  astronomy and cosmology.  If the current
paradigm,   where   the   initial   Gaussian  fluctuations   grew   by
gravitational  instability,  is  correct, the  two--point  correlation
function of  galaxies is  directly related to  the initial  mass power
spectrum.  While  the role of  the two--point correlation  function is
central, estimators for extracting it from a set of spatial points are
confusingly  abundant in the  literature. We  have collected  the nine
most important  forms used in the  area of mathematics  and astronomy. 
The difference between them lies  mainly in their respective method of
edge correction.

The  multitude of  choices might  appear confusing  to  the practicing
observational astronomer.  The  reason is partly due to  the lack of a
clear criterion  to distinguish between the  estimators.  For instance
one estimator could have smaller variance under certain circumstances,
but it could have a  bias. Therefore, {\em before} doing any numerical
experiments,  we  agreed upon  the  method  of  ranking the  different
estimators.  The  cumulative probability distribution  of the measured
value lying within a certain  tolerance of the ``true'' value is going
beyond the concepts  of bias or variance, and  even takes into account
any   non--Gaussian  behavior   of  the   statistics.   This   is  the
mathematical formulation of the simple idea that an estimator which is
more likely to give values closer to the truth is better.
After  the above  criterion  was  agreed, the  plan  to elucidate  the
confusion was clear.  Collect  the different forms of estimators (next
section), perform  a numerical experiment  in several subsamples  of a
large simulation  (\S 3), determining  the cumulative probability
of  measuring  values close  to  the  true  one, thereby  ranking  the
different estimators (\S 4).

\section{The estimators}

Astrophysical studies favor estimators  based on counting pairs, while
most  of  the  mathematical  research  is focused  on  geometric  edge
correction (second subsection). The following subsections collect nine
of the most successful and widespread recipes from both genres.

\subsection{Pairwise estimators}

Following   {}\citet{szapudi:new}  (hereafter   SS),  let   us  define
the pair--counts  with  a function $\Phi$ symmetric in its
arguments
\begin{equation}
    P_{DR}(r) = \sum_{\bx\in D}\sum_{\by\in R} \Phi_r(\bx,\by) .
\end{equation}

The summation runs over coordinates of  points in the data set $D$ and
points in  the set $R$  of randomly distributed  points, respectively.
This {\em  letter} considers the  two point correlation  function, for
which  the  appropriate  definition  is  $\Phi_r(x,y)=[r\le  d(x,y)\le
r+\Delta]$, where  $d(x,y)$ is the  separation of the two  points, and
[{\it  condition}]   equals  $1$  when  {\it   condition}  holds,  $0$
otherwise.  $P_{DD}$  and $P_{RR}$  are defined analogously,  with $x$
and $y$  taken entirely  from the data  and random samples,  under the
restriction that  $x \ne y$.   Let us introduce the  normalized counts
$DD(r)=P_{DD}(r)/(N(N-1))$,         $DR(r)=P_{DR}(r)/(N         N_R)$,
$RR(r)=P_{RR}(r)/(N_R(N_R-1))$,  with $N$  and $N_R$  being  the total
number of data and random points in the survey volume.
With  the  above preparation  the  pairwise  estimators  used in  what
follows  are the  natural estimator  $\widehat{\xi}_{\rm N}$,  and the
estimators  due to  {}\citet{davis:surveyV}  $\widehat{\xi}_{\rm DP}$,
{}\citet{hewett:estimation}          $\widehat{\xi}_{\rm         He}$,
{}\citet{hamilton:towards}      $\widehat{\xi}_{\rm      Ha}$,     and
{}\citet{landy:bias} (hereafter LS) $\widehat{\xi}_{\rm LS}$:
\begin{equation}
\widehat{\xi}_{\rm N} = \frac{DD}{RR}-1, \quad
\widehat{\xi}_{\rm DP} = \frac{DD}{DR}-1, \quad
\widehat{\xi}_{\rm He} = \frac{DD-DR}{RR}, \quad
\widehat{\xi}_{\rm Ha} = \frac{DD\ RR}{DR^2}-1, \quad
\widehat{\xi}_{\rm LS} = \frac{DD-2DR+RR}{RR}.
\end{equation}
Note that Hewett's estimator could  be rendered equivalent with the LS
estimator   if  the   original  asymmetric   definition  of   $DR$  is
symmetrized;  the  version we  use  is  the  one consistent  with  the
notation laid out above. In the  case of an angular survey, an optimal
weighting scheme can be adapted  to any of the above estimators (e.g.,
{}\citealt{colombi:effects}).  This  is inversely proportional  to the
errors   expected  at  a   particular  pair   separation,  essentially
equivalent to the {}\citet{feldman:power} weight.

\subsection{Geometric Estimators}

Alternative estimates of the  two--point correlation function from $N$
data points $\bx\in D$ inside a  sample window $\CW$ may be written in
the form
\begin{equation}
\widehat{\xi}(r) + 1 = \frac{|\CW|}{N(N-1)} 
\sum_{\bx\in D} \sum_{\by\in D}
\frac{\Phi_r(\bx,\by)}{4\pi r^2\Delta}\ \omega(\bx,\by) .
\end{equation}
$|\CW|$ is the  volume of the sample window and  the sum is restricted
to  pairs of  different  points $\bx\ne\by$.   For  a suitably  chosen
weight function $\omega(\bx,\by)$  these edge corrected estimators are
approximately unbiased.
Such     weights    are    the     {}\citet{ripley:second-order}    --
{}\citet{rivolo:two-point}     weight     $\omega_{\rm    R}$,     the
{}\citet{ohser:onsecond} -- {}\citet{fiksel:edge} weight $\omega_{\rm
F}$, and the {}\citet{ohser:estimators} weight $\omega_{\rm O}$.
\begin{equation}
\omega_{\rm R}(\bx,\by) = 
\frac{4\pi r^2}{\mbox{area}(\partial\CB_{r}(\bx)\cap\CW)}, \quad
\omega_{\rm F}(\bx,\by) = 
\frac{|\CW|}{\gamma_{\CW}(\bx-\by)},\quad
\omega_{\rm O}(\bx,\by) = 
\frac{|\CW|}{\overline{\gamma_{\CW}}(|\bx-\by|)}
\end{equation}
where  $\mbox{area}(\partial\CB_{r}(\bx)\cap\CW)$ is  the  fraction of
the surface area of  the sphere $\CB_r(\bx)$ with radius $r=|\bx-\by|$
around      $\bx$      inside      $\CW$,     the      set--covariance
$\gamma_\CW(\bz)=|\CW\cap\CW_\bz|$ is  the volume of  the intersection
of the original sample $\CW$  with the set $\CW_\bz$ shifted by $\bz$,
and $\overline{\gamma_{\CW}}(r)$ is the isotropized set--covariance.
We   will    consider   the   estimators    $\widehat{\xi}_{\rm   R}$,
$\widehat{\xi}_{\rm  F}$, and $\widehat{\xi}_{\rm  O}$ based  on these
weights.  A detailed  description of these estimators may  be found in
{}\citet{stoyan:stochgeom} and {}\citet{kerscher:twopoint}.
The Minus  or reduced sample estimator, employing  no weighting scheme
at  all, may  be  obtained by  looking  only at  the $N^{(r)}$  points
$D^{(r)}$ which are further than $r$ from the boundaries of $\CW$:
\begin{equation}
\widehat{\xi}_{\rm M}(r) + 1 = 
\frac{|\CW|}{N}\frac{1}{N^{(r)}} \sum_{\bx\in D^{(r)}} \sum_{\by\in D} 
\frac{\Phi_r(\bx,\by)}{4\pi r^2\Delta}
\end{equation}
Estimators of this type are used by {}\citet{labini:scale}.

It can be shown that  the natural estimator $\widehat{\xi}_{\rm N}$ is
the  Monte--Carlo version of  the Ohser  estimator $\widehat{\xi}_{\rm
O}$.   Similarly, the geometric  counterparts of  the LS  and Hamilton
estimator   may  be  constructed   {}\citep{kerscher:twopoint}.   This
allowed us  to cross-check our  programs. Focusing on  improved number
density  estimation, {}\citet{stoyan:improving}  arrived  also at  the
geometrical version of the Hamilton estimator.

\section{The comparison}

To compare these estimators for typical cosmological situations we use
the cluster  catalogue generated from  the $\Lambda$CDM Hubble--volume
simulation  {}\citep{colberg:clusters}.  In  order to  investigate the
effects of shape,  clustering, and the amount of  random data used, we
have always varied  one parameter at a time,  starting from a fiducial
sample.   Rectangular subsamples  were extracted  exhausting  the full
simulation cube:  the fiducial cubic  subsamples {\bf C},  slices {\bf
S}, pencil beams {\bf P}, and cubic samples with cutout holes {\bf H},
all  with approximately  the same  volume and  with  approximately 430
clusters  each.
Cutout  holes  around  bright  stars,  etc.  arise  naturally  in  all
realistic  surveys. The  pattern  of  holes used  for  this study  was
directly  mapped from a  $19^{\circ} \times  19^{\circ}$ patch  of the
EDSGC survey  to one of the  faces of the  simulation sub--cube.  Then
the holes were continued across the subsample, parallel with the sides
corresponding to a distant  observer approximation.  The physical size
of the holes  roughly corresponded to a redshift survey  at a depth of
about $300\hMpc$.  All these point  sets are ``fully sampled'' and may
be considered as volume limited samples.

In addition,  Poisson samples {\bf N}, i.e.\  without clustering, were
generated.  All calculations employed $N_R=100$k random points for the
pairwise estimators, unless otherwise  noted.  This was sufficient for
all indicators to converge.  The calculations were repeated for sample
{\bf C} with  $N_R=1$k and $N_R=10$k random points,  denoted with {\bf
R1}  and  {\bf  R10},   respectively,  to  investigate  the  speed  of
convergence  of the different  estimators with  respect to  the random
point  density.  The  parameters  for the  samples  are summarized  in
Table~\ref{table:samples}.
\placetable{table:samples}

The two--point correlation function $\xi_{\rm per}$ extracted from all
clusters  inside the $3\hGpc$  cube provided  our reference  or ``true
value''.  Since the simulation  was carried out with periodic boundary
conditions, the  cluster distribution is also  periodic, therefore the
torus boundary correction is exact {}\citep{ripley:spatial}.

The  nine  estimators defined  above  for  the two--point  correlation
function were  determined from  each of the  $n_S$ subsamples.   For a
given     radial    bin    $r$     we    computed     the    deviation
$|\widehat{\xi}_{\star}(r)-\xi_{\rm   per}(r)|$   of   the   estimated
two--point  correlation function  $\widehat{\xi}_{\star}(r)$  from the
reference $\xi_{\rm per}(r)$.
The empirical  distribution of these deviations  provides an objective
basis for the comparison of  the utility of the estimators.  The large
number of samples enabled  the numerical estimation of the probability
$P(|\widehat{\xi}_{\star}-\xi_{\rm   per}|<d)$   that  the   deviation
$|\widehat{\xi}_{\star}-\xi_{\rm  per}|$ is  smaller than  a tolerance
$d$.  The larger this probability,  the more likely that the estimator
will be within the predetermined range from one sample.  In general it
could happen that the rank of two estimators reverses as the tolerance
varies,  but as  will be  shown  in the  next section,  this is  quite
atypical.
This  procedure  is  more  general  than  considering  only  bias  and
variance,  which   yields  only  a  full  description   if  the  above
distribution is the integral of a Gaussian.  Note that a small bias is
negligible  for  practical  purposes  if the  variance  dominates  the
distribution of the deviations. It is worthwhile to note that Gaussian
assumption yields  a surprisingly  good description of  the deviations
$|\widehat{\xi}_{\star}(r)-\xi_{\rm per}(r)|$.   For estimators of the
closely  related  product   density,  asymptotic  Gaussianity  of  the
deviations was proven by {}\citet{heinrich:asymptotic}.

\begin{figure}
\figurenum{1} 
\epsscale{0.9}
\plotone{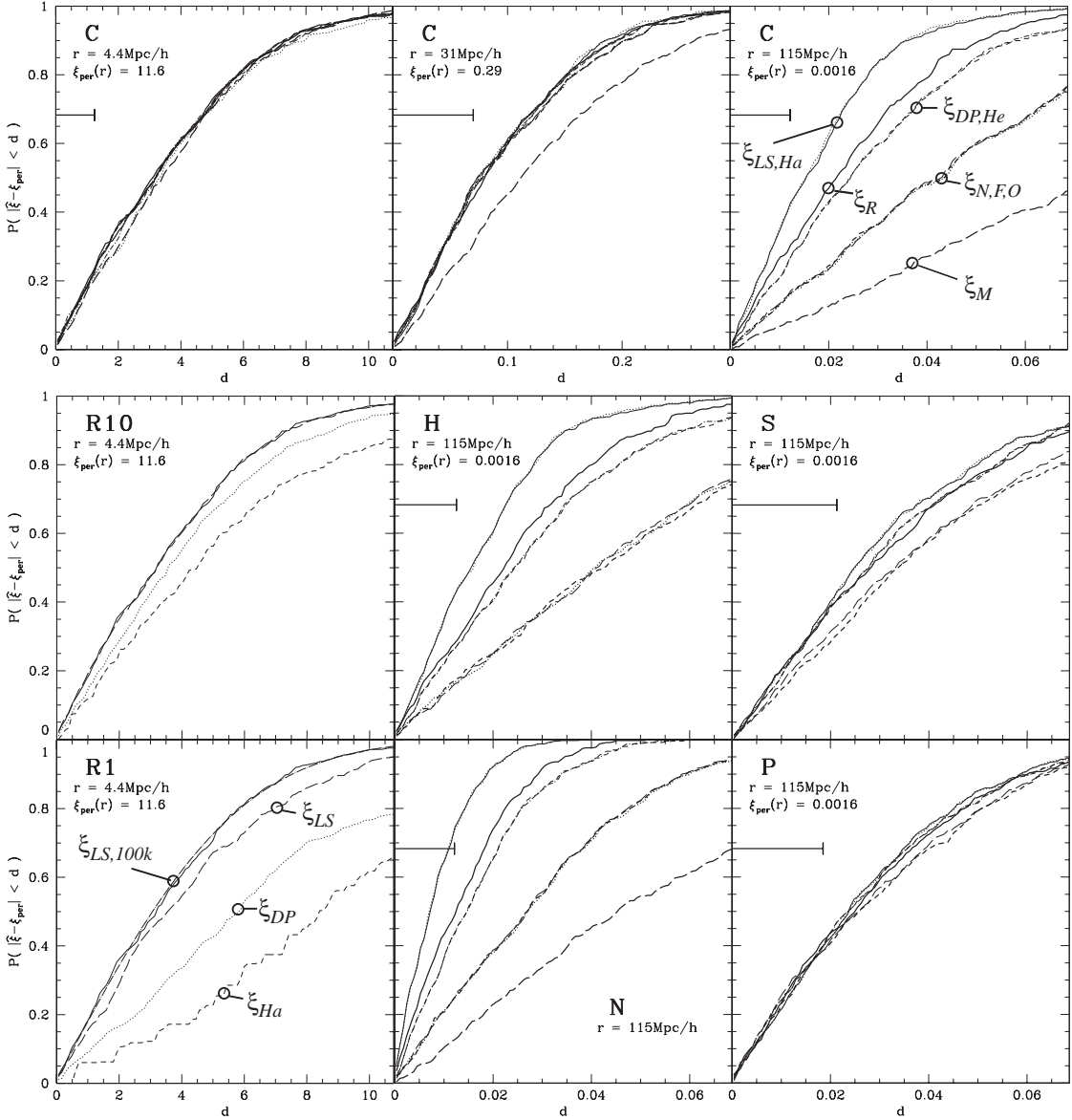} 
\figcaption[fig.eps]{   The    cumulative   probability   distribution
  $P(|\widehat{\xi}_{\star}(r)-\xi_{\rm per}(r)|<d)$ of the deviations
  $d$ are shown  for several samples and radii $r$.   In the plots for
  the samples {\bf C, S, P, H, N} we use the following symbols for the
  deviations   of   the   estimators:   Natural  (long   dashed),   DP
  (dotted--dashed),  Hewett  (short  dashed),  Hamilton  (dotted),  LS
  (solid), Rivolo (solid, thick), Fiksel (dotted, thick), Ohser (short
  dashed, thick),  and Minus (long  dashed, thick).  (On  large scales
  neither the Minus nor the  Fiksel are applicable in the samples {\bf
  S} and {\bf P} and consequently no results for them are shown.)  The
  horizontal lines  starting at 0.683 marks the  value of $\sigma_{\rm
  LS}$ for a Poisson  process according to Eq.~(\ref{eq:varLS}) in the
  geometry considered.   In the plots for  {\bf R1} and  {\bf R10} the
  solid line marks the result  for the LS estimator using 100k points.
  The  estimators using  1k  and 10k  random  points respectively  are
  marked in  the following way: DP (dotted),  Hamilton (short dashed),
  LS (long dashed).  For better  visibility we lay a Gaussian distance
  distribution (dashed--dotted)  over the result for  the LS estimator
  only in the plot for {\bf R1}.  A similar perfect agreement would be
  obtained in the other cases.
\label{fig:distribution}}
\end{figure}

\placefigure{fig:distribution}
Fig.~\ref{fig:distribution}        shows        the       distribution
$P(|\widehat{\xi}_{\star}(r)-\xi_{\rm  per}(r)|<d)$  for  the  samples
described  in  Table~\ref{table:samples}.   Three typical  scales  are
displayed   to  illustrate  the   general  behavior.    The  principal
conclusions to be drawn are the following:\\

{\em  Small  scales  ($r=4.4\hMpc$):}   the  effect  of  any  boundary
correction  scheme  becomes  negligible,  and  as  expected,  all  the
estimators exhibit  nearly identical behavior.   The same is  true for
the  samples  {\bf S},  {\bf  P},  {\bf N},  and  {\bf  H} not  shown.
However, some of  the estimators are more sensitive  to the density of
random  points, especially  the  Hamilton estimator,  followed by  the
Davis--Peebles estimator.  They show stronger deviations  for the {\bf
R1} and  {\bf R10} samples due to  the poor sampling of  the $DR$ term
(see  also {}\citealt{ponsborderia:comparing}).  This  effect persists
on large scales as well.

{\em Intermediate scales ($r=31\hMpc$):}  similar to the small scales.
The  Minus  estimator  shows  stronger deviation  becoming  even  more
pronounced for  the {\bf S}, {\bf  P}, and {\bf H}  samples, since the
effective remaining volume decreases.

{\em  Large  scales  ($r=115\hMpc$):}  edge corrections  are  becoming
important,  and  the estimators  exhibit  clear  differences in  their
distributions of the deviations for  the samples {\bf C}, and {\bf N}.
For  a  given  probability,  the  Minus estimator  shows  the  largest
deviations,  followed by  the  Natural, Fiksel  and Ohser  estimators.
Significantly smaller  deviations are obtained  for the Davis--Peebles
and  Hewett estimators,  and  yet smaller  for  the Rivolo  estimator.
Finally, the Hamilton and LS estimator display the smallest deviations
and  thus the  best  edge correction.   The  two latter  distributions
nearly  overlap.  The  above conclusions  are robust  and  only weakly
influenced by the  presence of cutout holes, as seen  from the {\bf H}
sample.  The geometry  of the subsamples has a  non--trivial effect on
the distributions.  While the deviations  are increased in the {\bf S}
and  {\bf  P} samples,  the  differences  between  the estimators  are
reduced in  the {\bf S}, becoming  negligible in the  {\bf P} samples.
In both cases,  the Minus estimator is not usable  any more, since the
$N^{(r)}$ equal zero, whereas the  Fiksel estimator is biased for such
geometries on  large scales (this is  implicitly shown in  the work of
{}\citet{ohser:estimators}).

Following {}\citet{szapudi:variance} the  variance of the LS estimator
may    be   calculated    for    a   Poisson    process:
\begin{equation}
\label{eq:varLS}
\sigma^2_{\rm LS}(r) = \frac{2}{V_\Delta(r)\ \overline{\rho}^2} ,
\end{equation}
with   $V_\Delta(r)=\int_\CW\rmd^3x\int_\CW\rmd^3y\  \Phi_r(\bx,\by)$.
The $\sigma_{\rm  LS}$ calculated for  the considered samples  is also
shown   in   Fig.~\ref{fig:distribution}   to  illustrate   how   much
discreteness effects contribute to the distribution of the deviations.
For our  choice of  sample parameters, the  discreteness contribution,
i.e.\  the deviation  of  a corresponding  Poisson  sample, is  always
within  a  factor of  few  of  other  important contributions  to  the
variance, such  as finite  volume and edge  effects.  In  general, the
ratio  of discreteness  effects  to  the full  variance  depends in  a
complicated  non--linear fashion  on  the number  of  clusters in  the
sample,  the  shape  of  the  survey, integrals  over  the  two--point
correlation function  and its square, and the  three-- and four--point
correlation  functions (see  {}\citealt{szapudi:unpub2} for  the exact
calculation).
Varying  the  side  length  of  the  cubic  samples  from  $300\hMpc$,
$375\hMpc$, $600\hMpc$  to $1\hGpc$ we  explored the influence  of the
size of the sample on the discreteness effects.  Still, the rank order
of the estimators stayed invariant.

\section{Summary and Conclusion}

For a  sample with  222,052 clusters extracted  from the  Virgo Hubble
volume  simulation  a reference  two--point  correlation function  was
determined.   Within over  500 subsamples  several estimators  for the
two--point  correlation  function   were  employed,  and  the  results
compared with the reference value.
On small  scales all the  estimators are comparable.  On  large scales
the LS  and the Hamilton estimator significantly  outperform the rest,
showing the  smallest deviations for  a given cumulative  probability. 
While the  two estimators yield almost identical  results for infinite
number of  random points, the Hamilton estimator  is considerably more
sensitive to the number of random points employed than the LS version.
From a practical point of view the LS estimator is thus preferable.
The rest of  the estimators can be divided  into three categories: The
first runner--ups  are the estimators from  Rivolo, Davis--Peebles and
Hewett,  but already  with a  significantly increased  variance.  Even
larger  deviations are  present  for the  Natural,  Fiksel, and  Ohser
recipes.  The Minus estimator has the largest deviation.
Although it was shown that for special point processes both the LS and
Hamilton estimator are biased {}\citep{kerscher:twopoint}, the present
numerical  experiment demonstrates  that  this is  irrelevant for  the
realistic  galaxy and  cluster point  processes,  as the  bias has  an
insignificant effect on the distribution of deviations.
{}\citet{ponsborderia:comparing} did  not recommend one  estimator for
all cases.  In contrast, through our extensive numerical treatment the
LS estimator emerges as a clear recommendation.

The above  considerations apply to  volume limited samples.   When the
correlation function is estimated  directly from a flux limited sample
with    an    appropriate     minimum    variance    pair    weighting
({}\citet{feldman:power}), the Hamilton estimator has the advantage of
being independent of the the normalization of the selection function.

The differences  between the estimators  become smaller for  the slice
and insignificant for the pencil beam samples.  At first sight this is
counter intuitive:  the difference  between the estimators  is largely
due to edge corrections, and  less compact surveys obviously have more
edges.  For  the large  scale considered, {\bf  S} and {\bf  P} become
essentially    two--    and    one--dimensional   and    the    weight
$\omega(\bx,\by)\approx\overline{\omega}(r)$ is equal  for most of the
pairs  separated   by  $r$.    Since  the  geometric   estimators  are
approximately   unbiased   they   employ   mainly  the   same   weight
$\overline{\omega}(r)$ on large scales, and consequently show the same
distribution  of deviations. This  argument also  applies to  the pair
estimators,  since they  may  be  written in  terms  of these  weights
{}\citep{kerscher:twopoint}.

All the  above numerical investigations are intimately  related to the
problem  of   calculating  the  expected  errors   on  estimators  for
correlation functions.   To include  all contributions, such  as edge,
discreteness,   and    finite   volume   effects,    the   method   by
{}\citet{colombi:large},                      {}\citet{szapudi:cosmic},
{}\citet{colombi:effects}, {}\citet{szapudi:unpub1} has to be extended
for  the  two point  correlation  function.   Such  a calculation  was
performed       by      {}\citet{szapudi:unpub2},       (see      also
{}\citealt{stoyan:variance},     {}\citealt{bernstein:variance}    and
{}\citealt{hamilton:towards}  for approximations)  and should  be used
for ab initio error calculations.

It is worth to mention, that one of the most widely used method in the
literature,  ``bootstrap'',  is based  on  a  misunderstanding of  the
concept.  For  bootstrap in spatial  statistics, a {\em  whole} sample
takes the role of one point in the original bootstrap procedure.  This
means that replicas of the original surveys would be needed to fulfill
the  promise of  the bootstrap  method.  Choosing  {\em points}  (i.e. 
individual  galaxies, clusters,  etc.)  randomly  from one  sample, as
usually  done, yields  a  variance  with no  obvious  relation to  the
variance sought (see also {}\citealt{snethlage:bootstrap}).

The role of the random samples is to represent the shape of the survey
in a Monte--Carlo  fashion.  A practical alternative is  to put a fine
grid on  the survey and calculate  the quantities $DD,  DW$, and $WW$,
where $D$  now represents bin--counts  and $W$ the  indicator function
taking the value one for pixels inside the survey, and zero otherwise.
According to SS,  all the above estimators have  an analogous ``grid''
version (see also  {}\citealt{hamilton:towards}) which can be obtained
formally  by  the substitution  $R\rightarrow  W$.   In practice  grid
estimators can  be more efficient than  pair counts, and  except for a
slight  perturbation of  the  pair separation  bins,  they both  yield
almost identical results for scales larger than a few pixel size.

The usual way  of estimating the power spectrum,  using a folding with
the Fourier transform of the sample geometry is equivalent to the grid
version of  the LS estimator.   Hence, such a power  spectrum analysis
extracts the same amount of  information from the data as the analysis
with the two--point correlation function using the grid version of the
LS  estimator.  The  results  are  only displayed  with  respect to  a
different  basis.   Similarly,  Karhunen--Lo\`e{}ve  (KL)  modes  form
another set of basis functions {}\citep{vogeley:eigenmode}.
The  uncorrelated power  spectrum  {}\citep{hamilton:uncorrelated} and
the  KL modes  are the  methods of  choice for  cosmological parameter
estimation.   The KL  modes allow  for a  well--defined  cut--off, and
therefore  reduce  the computational  needs  in  a maximum  likelihood
analysis.   However, geometrical  features of  the galaxy  and cluster
distribution directly  show up in two--point  correlation function and
may  be interpreted easily.   Each bin  of the  two--point correlation
function contains  direct information on pairs separated  by a certain
distance,  an intuitively simple  concept more  suitable to  study and
control  (expected or  unexpected) systematics  (geometry, luminosity,
galaxy  properties, biases)  than any  other representation.   In this
sense the  correlation function is  a tool complementary to  the power
spectrum.

\section*{Acknowledgments}

We   are   grateful   to   the   Virgo   Supercomputing   Consortium\\
{}\url{http://star-www.dur.ac.uk/$^\sim$frazerp/virgo/virgo.html}, who
made the Hubble volume simulation data available for our project.  The
simulation was  performed on  the T3E at  the Computing Centre  of the
Max-Planck Society  in Garching.  We  would like to thank  Simon White
and   the  referee   Andrew  Hamilton   for  useful   suggestions  and
discussions. MK would like to thank Claus Beisbart and Dietrich Stoyan
for  interesting and  helpful discussions.   IS was  supported  by the
PPARC  rolling  grant for  Extragalactic  Astronomy  and Cosmology  at
Durham   while  there.    MK  acknowledges   support  from   the  {\em
Sonderforschungsbereich f{\"u}r  Astroteilchenphysik SFB~375 der DFG}.
AS has been supported by NSF AST9802980 and NASA LTSA NAG653503.



\begin{table}
\caption{ The description of the  samples. $L_{xy}$, and $L_z$ are the
side--lengths  of the  rectangular  samples, $n_S$  is  the number  of
samples exhausting  the periodic cube with 222,052  clusters in total,
$\overline{N}$  the mean number  of clusters  inside the  samples, and
$N_R$ the number of random points used.
\label{table:samples}
}
\begin{center}
\begin{tabular}{l||r|r|r|r|r|r}
& {\bf C} & {\bf S} & {\bf P}  & {\bf H} & {\bf N} &  {\bf R1/R10}\\[1ex]
$L_{xy}$ [Mpc/h] & 375 & 1000 & 130  & 375 & 375 & 375 \\
$L_z$\ \ [Mpc/h]    & 375 & 52.6 & 3000 & 375 & 375 & 375 \\
$n_S$          & 512 & 513  & 529  & 512 & 512 & 512 \\ 
$\overline{N}$ & 434 & 433  & 420  & 434 & 434 & 434 \\ 
$N_R$          & 100k & 100k  & 100k  & 100k & 100k & 1k/10k \\ 
\end{tabular}
\end{center}
\end{table}

\end{document}